Scared into Action: How Partisanship and Fear are Associated with Reactions to Public Health

Directives

Mike Lindow[1]*, David DeFranza[1], Arul Mishra[1], Himanshu Mishra[1]

[1] University of Utah, David Eccles School of Business

* Corresponding Author

1655 Campus Center Dr., Salt Lake City, UT 84112

Mike.Lindow@utah.edu

Abstract

Differences in political ideology are increasingly appearing as an impediment to successful bipartisan communication from local leadership. For example, recent empirical findings have shown that conservatives are less likely to adhere to COVID-19 health directives. This behavior is in direct contradiction to past research which indicates that conservatives are more rule abiding, prefer to avoid loss, and are more prevention-motivated than liberals. We reconcile this disconnect between recent empirical findings and past research by using insights gathered from press releases, millions of tweets, and mobility data capturing local movement in retail, grocery, workplace, parks, and transit domains during COVID-19 shelter-in-place orders. We find that conservatives adhere to health directives when they express more fear of the virus. In order to better understand this phenomenon, we analyze both official and citizen communications and find that press releases from local and federal government, along with the number of confirmed COVID-19 cases, lead to an increase in expressions of fear on Twitter.

History has shown that crisis unites communities; people move beyond everyday differences to overcome common adversity. During Hurricane Harvey, for instance, citizens of Houston spontaneously volunteered, risking their own wellbeing, to help others from cars and into boats (Adams, 2017). During the historic heat wave in Chicago in 1995, older residents in urban areas were sheltered by others in the community, leading to dramatically better outcomes than their more isolated suburban counterparts (Klinenberg, 2015). During Hurricane Katrina, members of the community gathered supplies from abandoned stores to support those stranded in the Superdome (Glass, 2005). Indeed, examples abound from the London Blitz (Solnit, 2010) to the September 11, 2001 terrorist attacks and the uncertainty that followed (Steffen & Fothergill, 2009; Umbreit, Lewis, & Burns, 2003), which demonstrate that communities ignore differences and rally to support one another during a disaster (Haraoka, Ojima, Murata, & Hayasaka, 2012; Storr & Haeffele-Balch, 2012). Given this evidence, one would predict that in the face of a pandemic like COVID-19, citizens across the country would put aside social and political differences to cooperate in order to save lives. Instead, we have seen that a national crisis which typically would engender unity has been politicized. This is congruent with recent events like the politicization of historically non-partisan government agencies (Jamison, 2015; Mulgan, 2007; Pace, Sullivan, & Sakama, 2017; Richardson, 2017), the partisan divide on Ebola preparedness (Nyhan, 2014), and is in line with the recent polarization of American politics in general (Pew, 2014). Therefore, it is not surprising that conservatives and liberals have responded to the COVID-19 pandemic differently. There is growing evidence suggesting conservatives are more likely to question shelter-in-place directives, see the pandemic as a liberal hoax, refuse to wear masks or social distance, and protest against public health directives in general (Van Green & Tyson, 2020). Recent research specific to political polarization has also found that conservatives are less likely to restrict their movements during shelter-in-place directives (van Holm, Monaghan, Shahar, Messina, & Surprenant, 2020), less likely to social distance (Painter & Qiu, 2020), and less likely to search for information about COVID-19 (Barrios & Hochberg, 2020).

In other words, both evidence from research and reports from the media suggest that conservatives demonstrate less compliance to COVID-19 directives than liberals. In the absence of a unified front against the virus, it is more difficult to curb its spread. Therefore, in this research we move beyond examining whether conservatives are less likely to adhere to health directives and instead explore what can motivate them to do so. Empirically, we utilized large scale mobility data to understand changes (or lack thereof) in behavior, analyzed millions of tweets using word embedding models to measure expressed fear, and sought ways to constructively use fear appeals using the XGBoost algorithm. We focused on tweets because past research has shown social media is pivotal for the dissemination and consumption of emotional political content (Jost, Bonneau, Langer, Metzger, & Nagler, 2018) and emotional language on social media increases political message diffusion (Brady, Will, Jost, Tucker, & van Bavel, 2017).

**What can Motivate Conservatives to Adhere?**

In order to determine what could motivate people to put aside partisanship and confront the pandemic in a united manner, we look to a long stream of research considering political ideology. Recent work that has shown less adherence and increased skepticism of health directives by conservatives (Iyengar & Massey, 2019) is surprising because it runs counter to past research, which has demonstrated that conservatives are more likely to be rule abiding, risk averse and prevention focused (Jost, Glaser, Kruglanski, & Sulloway, 2003b; Jost, Napier, Thorisdottir, Gosling, Palfai, & Ostafin, 2007; Sales, 1973). In times of crisis, conservatives are more likely than liberals to seek safety (Sales, 1973; Thorisdottir & Jost, 2011).  But given the current tug-of-war between political rhetoric and health risk (Barrios & Hochberg, 2020; Cruwys Stevens, & Greenaway, 2020; Panagopoulos, Kerr, & van der Linden, 2020; Rothgerber, Wilson, Whaley, Rosenfeld, & Humphrey, 2020) research has indicated that conservatives rely more on political identity (Allcott, Boxell, Conway, Gentzkow, & Thaler, 2020; Kushner, Goodman, & Pepinsky, 2020; Painter & Qiu, 2020; van Holm et al., 2020), which could be influenced or exacerbated by conservative officials downplaying the health risks of the coronavirus (Bursztyn Rao, Roth, & Yanagizawa-Drott, 2020; Peters, 2020). One specific attribute commonly associated with a conservative mindset may explain why some have diminished or ignored the importance of COVID-19 health directives and what could motivate these individuals to view the pandemic more seriously. This attribute is fear.

Research on political identity has demonstrated that conservatives, more than liberals, display a greater reaction to fear in response to threats in the environment (Block & Block, 2006; Jost et al., 2003a; Jost, Glaser, Kruglanski, & Sulloway, 2003b; Oxley et al., 2008; Pliskin, Sheppes, & Halperin, 2015). That is, conservatives have a stronger reaction to threats and new experiences (Oxley et al., 2008) and express stronger emotional reactions to negative outcomes (Joel, Burton, & Plaks, 2014). Such a fear response could be driven by a greater need for control over the environment and greater impulse to reduce uncertainty (Jost et al., 2003a). Since conservatives generally respect authority and want the hierarchical structure to remain in place, they are more fearful of change to this structure (Adorno, Frenkel-Brunswik, Levinson, & Sandford, 1950; Jugert & Duckitt, 2009).

Herein lies the contradiction: If conservatives generally respond more to fear why are they complying less with health directives meant to curb the spread of a deadly virus? One reason for the observed lack of adherence to COVID-19 directives could be a greater reliance on political rhetoric that portrays COVID-19 as less harmful or innocuous (Panagopoulos et al., 2020). Specifically, in the case of COVID-19, researchers have observed polarization and partisan tweets in discussions of the pandemic (Green, Edgerton, Naftel, Shoub, & Cranmer, 2020). This research found that on Twitter, Republican leaders focused more on business and China while Democratic leaders focused on health and aid; such party line messaging can affect attitudes and subsequent behaviors of individuals who strongly identify with a specific political ideology (Grossman et al., 2020).

These two opposing findings can be reconciled if conservatives who do experience fear are more likely to adhere to health directives and display less mobility under shelter-in-place directives as compared to conservatives who do not experience fear. Such a pattern would not only explain the response of conservatives to COVID-19 directives and recommendations but would also suggest a path forward for policy makers intent on motivating greater adherence to health directives.

In order to test the influence of fear on adherence to health directives, we collected data across the 53 largest metropolitan areas in the United States over a 30-day period. During this period, we observed aggregate mobility both before and after shelter-in-place directives were imposed. Millions of tweets from metropolitan areas were used to assess fear. We used word embeddings to test our research proposition. Word embeddings are a Natural Language Processing (NLP) method in which distributional models of language capable of fine-grained semantic representation are learned. Such models have been applied extensively to infer meaning from text (Berger, Humphreys, Ludwig, Moe, & Netzer, 2020; Humphries & Wang, 2017). Word embeddings allowed us to measure the relative association between individual tweets about COVID-19 and a lexical representation of fear, then examine the three-way interaction between political ideology, issuance of shelter-in-place directives, and fear on subsequent changes in mobility.

## Political Ideology, Local Fear, and Adherence

Instead of relying on attitudes or intentions, we examine whether conservatives, as compared to liberals, increased or decreased movement in response to the shelter-in-place directives. We considered the most populous 53 metropolitan statistical areas with a population over one million, which included most of the 50 states and of which 28 implemented a shelter-in-place directive during our observation period. The data is a cross sectional time series, collected from the metropolitan areas over a 30-day period between February 29, 2020 and March 29, 2020 (N = 1590; see Appendix A for a detailed description of variables and data sources). February 29 was selected as the starting point for observation because the first COVID-related death in the United States was announced on this day (CDC, 2020). Data was collected for approximately one month (30 days) following the initial event. We next describe the data, including the intervention variable (i.e., shelter-in-place directives), predictor variables (political ideology and localized tweets), the outcome variable (i.e., mobility data indicating adherence), and control variables.

## Data

*Shelter-in-place intervention.* The imposition of the shelter-in-place directive provided a naturalistic intervention (Harrison & List, 2004; Meyer, 1995) to examine adherence to a health directive in a quasi-experimental approach amenable to generalized difference-in-differences analysis (Lechner, 2011; Wing, Simon, & Bello-Gomez, 2018). Shelter-in-place interventions varied on the local and state level across metropolitan areas. These occurrences were recorded during the observation period with binary variables indicating a local level or state level announcement of a shelter-in-place order. No orders were discontinued during this period. Of the 53 metropolitan areas, 28 were under a state issued shelter-in-place order during the 30-day period, while 25 were not.

*Political identity.* We measured the conservative or liberal identity of the metropolitan area using the voting record of the past five presidential elections within each county in the metropolitan area. Voting results were averaged for each county across the five presidential elections from 2000-2016 then aggregated to the level of the metropolitan area. Using the average of the last five presidential elections accounts for per-election variations in voting behavior and provides a more stable metric of political affiliation.

The political identity variable is a ratio of votes for a Republican candidate to votes for a Democrat candidate. Any metro area that has a value greater than one voted majority Republican while any metro area with a voting ratio of less than one voted majority Democrat over the past five presidential elections. Across the 53 metropolitan areas, 17 voted majority Republican and 36 voted majority Democrat.

*Localized tweets and measuring fear.* Over 2 million tweets were collected during the 30-day period. Tweets were collected if the message contained any occurrence (including but not limited to hashtags) of the words "coronavirus", "covid", "covid-19", or "sars-cov2" and the author's Twitter biography included location information indicating they were from one of the 53 metropolitan areas. For instance, any tweet with a corresponding location "San Francisco", "SF", "Oakland", or "Berkeley" was classified as a tweet from the San Francisco metropolitan area (see the Appendix B for a list of metropolitan areas and search terms).

A fear score for each tweet was calculated using distributed dictionary representations (DDR; Garten, Hoover, Johnson, Boghrati, & Iskiwitch, 2018) and GloVe pre-trained word embeddings (Pennington, Socher, & Manning, 2014). DDR has been used in the past to examine semantic language in Hurricane Sandy tweets (Hoover, Johnson, Boghrati, Graham, & Dehghani, 2018) and examine risk perceptions in open survey responses (Bhatia, 2019). Fear was represented using a validated dictionary of 26 synonyms that are most similar to the word fear based on cosine similarity (see Appendix C for a full list of words and detailed discussion of their selection; Nicolas, Bai, & Fiske, 2019). The results were also replicated with a smaller, more succinct dictionary (see Appendix C for the validation results). In order to compare the similarity of each tweet with the construct of fear, each word in the tweet was converted to a word vector. Tweet-level vectors were aggregated by taking the dot product of each word vector, divided by the sum of the Euclidean norm of each word vector. The result is a single 200-dimension vector representing each tweet. A similar process was followed to aggregate each word in the fear dictionary into a single 200-dimension vector. Subsequently, each tweet vector was compared to the fear vector using a metric of orientation, cosine similarity. In this way, it was possible to compare the similarity of each tweet to the construct of fear; greater cosine similarity indicated a higher association between the tweet and the fear dictionary.

*Google mobility measures.* To measure adherence, we used behavioral data provided by Google, which captures aggregate and anonymized mobility of Google users who have opted into the location history feature (Google, 2020). Data is decomposed into five subcategories: retail (restaurants, shopping centers, museums, libraries, etc.), grocery and pharmacy (markets, food warehouses, drugs stores, pharmacies, etc.), workplaces, parks (local and national parks, beaches, gardens, etc.), and transit stations (subway, bus, and train stations). The mobility metric is a percent change in the number of visits and length of stay compared to a baseline. The mobility baseline for each location is a five-week period from January 3, 2020 to February 6, 2020. For the purposes of the current analysis, Google mobility report data was collected for 30 days beginning February 29th and ending March 29th.The mobility data was reported at a county level by day and was then aggregated to the metropolitan area using a population-weighted mean.

*Control Variables.* Robustness checks of our focal analysis were conducted by controlling for location-specific and time-varying variables such as COVID-19 infection and death rates for each metropolitan area, state-issued or local school closure directives, time since directives were implemented, day of the week, average household income, and the poverty rate for each community. Control variables were aggregated to the level of the metropolitan area using the population weighted mean. See Appendix A for more information on control variables.

*Word Embedding Description.* Semantic analysis was conducted using tweets and DDR (Garten et al., 2018). First, dictionaries were developed consisting of a small set of words to represent each psychological construct. In doing so, an emphasis was placed on specificity rather than breadth (Dev et al., 2019). Next, tweets were normalized by making all text lowercase, removing numbers and punctuation, removing excess whitespace, lemmatizing words using WordNet and the Penn Treebank part of speech tags (Bird, Klein, & Loper, 2009; Marcus, Kim, Marcinkiewicz, MacIntyre, & Bies, 1994), and finally removing stop words and single character tokens (Symeonidis, Effrosynidis, & Arampatzis, 2018). For each word in each construct dictionary, a 200-dimension vector was obtained from publicly available pre-trained word embeddings (Pennington et al., 2014). These vectors were then aggregated within each construct by calculating the sum of the vectors divided by Euclidean norm of the vectors, essentially creating a single vector from the mean of the contributors. The same process was repeated for each tweet. The result was a single 200-dimension vector for each construct and a single 200-dimension vector for each tweet. Similarities were calculated using a metric of orientation, cosine similarity (Bolukbasi, Chang, Zou, Saligrama, & Kalai, 2016; Caliskan, Bryson, & Narayanan, 2017; Garten et al., 2018). For example, a tweet with a cosine similarity of one is most similar to the fear construct, and a tweet with a cosine similarity of -1 is most dissimilar to the fear construct. Finally, cosine similarity between individual tweets and the construct dictionary was aggregated by metropolitan area and by date, resulting in the average association between the fear construct and each metropolitan area, for each day.

**Model Specification**

Because of the nature of the cross sectional, grouped (by metropolitan area) time series data, we tested the effectiveness of ordinary least squares (OLS), fixed effects, and random effects models. After testing, we utilize random effects models (see Appendix E for diagnostic results; Hsiao 2014). Random effects models make an assumption that time invariant variables are uncorrelated with the time varying predictors, which allows an examination of time invariant variables (such as voting results) on the outcome variable (mobility). The three-way interaction model without control variables is explained in equation 1, where $m$ represents metropolitan area and $t$ represents time.

$$mobility_{m,t} = \mu_t + \beta SIP_{m,t} + \lambda voting\ results_m + \gamma lagged\ fear_{m,t}$$
$$+ \zeta \big(voting\ results_m * SIP_{m,t}\big) + \eta \big(voting\ results_m * lagged\ fear_{m,t}\big)$$
$$+ \delta \big(SIP_{m,t} * lagged\ fear_{m,t}\big)$$
$$+ \alpha \big(voting\ results_m * SIP_{m,t} * lagged\ fear_{m,t}\big) + \epsilon_{m,t}$$

Here $mobility_{m,t}$ is mobility level as reported by the Google mobility reports, $SIP_{m,t}$ is a shelter-in-place intervention dummy variable with value zero if shelter-in-place was not implemented at time $t$ and one if it was implemented, and $voting\ results_m$ is time invariant voting results for a metropolitan area. *Lagged fear$_{m,t}$* represented one-day lagged values of fear sentiment based on past work that has evaluated Twitter sentiment using a 1-day lag (Kaminski, 2016; Li, Situ, Gao, Yang, & Liu, 2017; Zhang, Wang, & Baoxin, 2013). Past work has shown that 75% of tweet replies are made within 17 minutes, and the mean flow of tweet conversations are six hours (Ye & Wu, 2010). Because a vast majority of Twitter users are passive, many tweets are ignored, implying that tweet impact relies heavily on being fresh in the Twitter feed (Romero, Galuba, Asur, & Huberman, 2011). For these reasons, the one-day lag of Twitter fear sentiment was used in the final models. $\epsilon_{m,t}$ is random disturbance and $\mu_t$ is the intercept which can vary over time. $\beta$, $\lambda$, $\gamma$, $\zeta$, $\eta$, $\delta$, and $\sigma$ are coefficients.

For each of the five mobility categories, a series of diagnostic tests were conducted to determine whether an OLS model, fixed effects model, or random effects model was preferred. An F-test determined if the fixed effects model was preferred over OLS regression, a Breush-Pagan LaGrange Multiplier test was used to determine whether a random effects model was preferred over OLS, and a Hausman test was used to determine whether the fixed effects model or the random effects model was preferred for each mobility category (see Appendix E for the diagnostic test results for each model). For all mobility categories, the random effects model provided the best fit.

## Results

*Political identity and shelter-in-place interactions.* The presence of a state shelter-in-place order is associated with a reduction in mobility. The relationship was significant for workplaces ($b$=-149.0849, $p$ = 0.0422) and parks ($b$ = -1364.6142, $p$ = 0.0166). However, the relationship was not significant for retail ($b$ = -104.5549, $p$ = 0.1850), grocery ($b$ = -153.8703, $p$ = 0.1039), and transit stations ($b$ = -176.8373, $p$ = 0.1241; see Table 1). This is in line with literature showing that shelter-in-place orders tend to reduce overall mobility (Dave et al., 2020).

Further, there is an interaction between political identity (i.e. a metropolitan area's Republican to Democrat voting ratio) and state shelter-in-place orders where in the presence of a shelter-in- place order, conservative metropolitan areas reduce mobility less. The interaction is significant (or marginally significant) for all mobility categories including retail ($b$ = 175.6505, $p$ = 0.0395), grocery ($b$ = 194.3915, $p$ = 0.0576), workplaces ($b$ = 174.3164, $p$ = 0.0281), parks ($b$ = 1799.8813, $p$ = 0.0035), and transit stations ($b$ = 227.9781, $p$ = 0.0668; see Table 1). This suggests that as the proportion of votes for Republican candidates increases residents in a metropolitan area are less likely to reduce mobility following a state shelter-in-place directive. Overall, this is consistent with recent literature that has found that in the presence of shelter-in-place orders more conservative areas struggle to reduce mobility (Allcott et al., 2020; Andersen, 2020; DeFranza, Lindow, Harrison, Mishra, & Mishra, 2020; Gollwitzer et al., 2020; van Holm et al., 2020).

*Three-way Interaction.* The three-way interaction between political identity, shelter-in-place orders, and lagged fear was significant (or marginally significant) for all mobility categories including retail ($b$ = -986.9036, $p$ = 0.0374), grocery ($b$ = -1065.4321, $p$ = 0.0610), workplaces ($b$ =-958.8522, $p$ = 0.0297), parks ($b$ = -9856.7920, $p$ = 0.0040), and transit stations ($b$ = -1228.0859, $p$ = 0.0757; see Table 1). In summary, the three-way interaction suggests that when a state issues a shelter-in-place order and fear is more salient among conservatives, residents are more likely to reduce their activity and mobility across all important dimensions of public life. The following five figures show the marginal effects plot of the relationship between mobility and the three-way interaction of political identity of a location, the presence of a shelter-in-place order, and the concentration of fear sentiment on Twitter for that location.

Our findings help reconcile past work, which has demonstrated that when conservatives experience fear, they are more likely than liberals to take preventive measures (Joel et al., 2014; Jost et al., 2007) and the current trend in which conservatives resist health directives (Irmak, Murdock, & Kanuri, 2020). Based on a large, ecologically valid sample, the results suggest that the relationship between one's political identity and adherence to health directives is more nuanced than what has been reported in past work. Specifically, as conservatives' expressions of fear of COVID-19 increase, so does local adherence to health directives. However, no such relationship is observed among majority liberal communities. At the same time, the data is the result of behavior during a calamitous and uncertain time. Future research should consider the consistency of such an effect under more staid conditions.

Given these findings, it is natural to wonder what the implications might be for policy makers. One obvious conclusion is that policy makers in majority conservative communities should increase the salience of fear when issuing directives. However, fear is a double-edged sword: on the one hand, instilling fear increases anxiety and leads to irrational decision making (Kligyte, Connelly, Thiel, & Davenport, 2013; Lee & Andrade, 2011; Mobbs & Kim, 2015; Wagner & Morisi, 2019) but on the other hand, allowing people to minimize the risk associated with COVID-19 is also harmful. Moreover, the pandemic has been politicized (Panagopoulos et al., 2020) and as such it is difficult to disseminate any related message free of partisan interpretation. Therefore, to identify a beneficial strategy that can increase caution, especially among conservatives, we next examine which predictors and covariates most influence fear. We specifically explore the role of press releases created by health policy makers in influencing the fear perceptions of the pandemic.

### The Effect of Press Releases on Fear

Past work in health policy has used government press releases to communicate information about health hazards to people (Fairbanks, Plowman, & Rawlins, 2007; Grossman, Kim, Rexer, and Thirumurthy, 2020; Lee & Basnyat, 2013; Levi & Stoker, 2000; Mayhew, 1974). Press releases are a common communication method used to inform a large number of citizens of a problem and the protocols for addressing the problem (e.g., what precautions need to be taken and how widespread the problem is or which groups of people might be most affected). Similar to past public health crises (e.g., disasters like Hurricane Katrina or outbreaks like that of swine flu or Ebola), for COVID-19, local governments have issued press releases to inform their constituents. Press releases are used to communicate information related to patient care, testing, and the provision of protective equipment, as well as providing guidelines tailored to constituents.

We focused on two types of press releases: local and national. Local COVID-19 related press releases were captured for the city (or county where applicable) of each of the 53 metropolitan areas. For example, for the San Francisco metro area, any press release from the city of San Francisco, Oakland, or Berkeley government websites that was related to COVID-19 was captured. Also, national press releases were captured in the same manor from official federal websites. Our data set of press releases is comprised of 166 national and 1232 local press releases (1,398 total) across metro areas during the 30-day period. As with tweets, we measured the sentiment of fear in press releases using the DDR method (Garten et al., 2018) with GloVe pretrained word vectors (Pennington et al., 2014). Fear sentiment for each press release was aggregated by metro area and again by day; for every metro area, there is a local and national press release fear score if a local or national press release was issued that day.

It is worth noting that not all metropolitan areas issued local press release every day or at all. There was significant heterogeneity across metropolitan areas with some areas issuing regular press releases and some areas, like Indianapolis, not issuing any press releases during the 30-day period. In total there were 347 observations where a city government issued a press release on a given day. Likewise, national press releases were not issued every day of the observation period. This heterogeneity reduced the number of data points from 1,590 to 291 (observations with both local and national press releases issued), rendering panel data analysis less useful or effective. Therefore, we employed gradient boosted decision trees (Chen & Guestrin, 2016; Friedman, 2001) to predict the level of fear in local tweets.

Gradient boosted decision trees employ a series of shallow, underfit decision trees (commonly referred to as stumps; Quinlan, 1986) to predict the outcome variable, under the assumption that many weak learners will achieve a probably approximately correct (PAC) result (Valiant, 1984). Each tree is evaluated based on the gradient of the error with respect to the prediction through a process known as functional gradient descent (Ruder, 2017). Improvements in prediction accuracy for subtrees with a steeper gradient lead to overall larger model improvements. Thus, gradient boosted trees result in the identification of variables that have the overall greatest influence on predictive accuracy. The method is amenable to analysis with relatively smaller sample sizes (Zhao & Duangsoithong, 2019), is not impacted by multicollinearity (Ding, Wang, Ma, & Li, 2016), and has been used in the past for prediction tasks involving social media popularity (Li et al., 2017) and tweets (Ong, Rahmanto, Suhartono, Nugroho, & Andangsari, 2017). In this analysis gradient boosted decision trees were implemented via extreme gradient boosting (commonly known as XGBoost; Chen & Guestrin, 2016).

## Variables, Model, and Training

The gradient boosted model predicts the next period local twitter fear score from the fear scores in the current period local and national press releases, COVID-19 cases and deaths, local political identity, presence of a shelter-in-place order, poverty levels, and median household income. We implemented a train-test split of 80/20, resulting in 235 training observations and 56 test observations. Each iteration of training was cross validated with 10 folds for the purposes of hyperparameter tuning.

The final tuned model employed 700 iterations, a learning rate of 0.025, a max tree depth of two, minimum child weight of 0.8, and gamma of zero. Each of the 700 iterations was also run with 10-fold cross validation. The tuned model subsamples half of the training set for each iteration (which prevents over fitting), while the base model by default uses the entire train set for every iteration (among other default tuning parameters). The tuned model outperforms the base default model by reducing root mean squared error (RMSE) 9% from 0.0061 to 0.0055 and increasing the $R^2$ value 34% from 0.3097 to 0.4153, essentially explaining 41% of the variation in local tweet fear.

## Results

Figure 6 shows the importance of each variable as a factor of how much it influenced the RMSE of the model. The horizontal axis of Figure 6 represents the prediction accuracy for each individual variable summed over each iteration of the boosted tree model. The results of examining individual variables shows that the most important factor in predicting local tweet fear for any given metropolitan area, on any given day, was the amount of national press release fear communicated the day before, followed by the number of confirmed cases for that area the day before. Other important variables include poverty levels, voting results, median household income, and the association of local press releases with fear. Of course, such an analysis only considers the overall valence of political communication and other features of rhetoric, such as argument quality.

## Probabilistic Analysis of Press Release Content

We have shown that fear plays an important role in adherence to health directives and that local and national press releases are important predictors of local fear. Next, we examine the content of the national and local press releases more closely to see what words or phrases are more likely to appear in press releases with a high (vs low) relative association with fear. Each word of the 166 national and 1232 local press releases (65,237 words total) were counted. A log odds ratio was calculated for each word showing the probability of each word occurring in press releases with high or low relative association with fear. Only words that occurred at least 10 times were considered. Any national or local press release that was above the median association with fear (0.5713) was considered high fear and low fear otherwise.

For each word, a log odds ratio was calculated which compares the log of the frequency of a given word in a high fear press release to the frequency of that word in a low fear press release (see equation 2). This gives a single value which represents the likelihood of a given word being in a high or low fear press release. If a word has a positive (negative) log odds ratio, it is more likely to be in a high (low) fear press release. A log odds ratio of zero means the word is equally likely to be in either high or low fear press releases.

**Results**

Figures 7 and 8 show the distribution of words in high or low fear press releases. Figures 9 and 10 show the most extreme words that occur at least 10 times in high or low fear press releases. High fear local press releases, compared to low fear, have more COVID-19 related words like people, covid, testing, flu, illness, cough, and chest. On the other hand, the low fear press releases tend to lack a focus on COVID-19. They contain words like fund, business, meeting, tax, and marketing. In local press releases the clear difference between low fear and high fear is a shift from a focus on COVID-19 to a focus on the procedural aspects of government.

National press releases tell a similar story, but with a national and economic focus. High fear national press releases focus on crises with words like people, public, testing, spread, and hospitals. The words that are most likely to appear in high fear national press releases include tests, cdc, testing, spread, and health indicating a focus on testing and healthcare similar to what is found in local high fear press releases.

## Discussion and Implications

An extensive body of literature suggests that conservatives are more inclined to comply with authority (Jost et al., 2003a; Jost et al., 2007; Sales, 1973; Thorisdottir & Jost, 2011) and are likely to take preventive measures when they are fearful of a situation (Joel et al., 2014). However, a more recent stream argues the opposite (Brehm & Brehm, 2013; Irmak et al., 2020) including evidence which suggests conservative non-adherence to COVID-19 health directives (Allcott et al., 2020; DeFranza et al., 2020; Gollwitzer et al., 2020; van Holm et al., 2020). We reconcile this contradiction by demonstrating that while it appears conservatives are not following health directives, this behavior is not completely monolithic. Rather than using attitudinal measures, we test our propositions using behavioral data as a response to shelter-in-place directives. We use Google mobility data, reported across the 53 most populous metropolitan areas, over a 30-day period and find that when members of majority conservative communities express a greater degree of fear of the pandemic, members of these communities are also more likely to shelter-in-place and reduce their mobility. Therefore, presence or absence of fear is an important determinant as to whether, at least in the United States, conservatives adhere to shelter-in-place directives.

We also examined how the expression of fear in government communication can be used in a helpful manner to increase caution and adherence. We utilized gradient boosted tree models to examine which variables have the greatest influence on predictions of fear expressed in tweets. The analysis indicates that an increased association with fear among national press releases as well as case counts are significant predictors of Twitter fear the following day. This additional analysis indicates that conservative areas should be regularly updated and informed via press releases and other official communications so that constituents understand the importance of taking precautions. Because discussions of COVID-19 have been partisan (Allcott et al., 2020), the role of local government press releases cannot and should not be ignored. If in these press releases, severity and caution about the virus are clarified, it could result in greater adherence. Statements diminishing risk, severity, and seriousness can cause significant harm in more conservative areas. In order to more deeply examine what language might be most effective in communicating the seriousness of COVID-19, we identified the words used most in press releases that expressed fear and caution as depicted in figures 9 and 10. This analysis could help health officials and policy makers to determine which words might be most effective in their communications. This result extends beyond the scope of COVID-19 and helps inform local government communication, specifically in conservative communities and for messages that may be important but unpopular.

Moreover, giving people, especially those in conservative areas, accurate and timely information (such as case counts) will help communities take appropriate precautions by engendering the appropriate attitude toward the objective risk. Our findings also demonstrate that the number of confirmed cases affect people's perception of the virus and makes them more cautious. Therefore, widespread testing and accurate reporting will help increase individual awareness and community safety. Restricted testing and underreporting case numbers (Krantz & Rao, 2020; Lau, Khosrawipour, Kocbach, Ichii, & Bania, 2020) could be detrimental to safe individual behavior and compliance with public health policy and directives. Moving beyond COVID-19 mitigation, this finding emphasizes the importance of timely, relevant, clear information causing an appropriate amount of motivational fear in more conservative communities.

**Figures and Tables**

Table1: *The relationship between political ideology, expressed fear, and the introduction of state issued shelter-in-place directives on mobility.*

| Variable | Retail $b$ (SE) | Grocery $b$ (SE) | Workplaces $b$ (SE) | Parks $b$ (SE) | Transit $b$ (SE) |
|---|---|---|---|---|---|
| Intercept | 41.3907 (44.5657) | 60.1436 (55.8535) | 57.2648 (41.6549) | 574.6642 (390.5824) | 23.4307 (36.4043) |
| Confirmed COVID cases | -0.0063*** (0.0013) | -0.0053*** (0.0015) | -0.0052*** (0.0012) | -0.0195** (0.0093) | -0.0101*** (0.0019) |
| Confirmed COVID deaths | 0.1053*** (0.0394) | 0.0942** (0.0473) | 0.0927** (0.0367) | 0.2373 (0.2846) | 0.1979*** (0.0566) |
| Day of the week | -0.0365 (0.0543) | -0.1041 (0.0651) | -0.0870* (0.0505) | -0.2193 (0.3922) | -0.1618** (0.0792) |
| State school closure | -0.2895 (0.6167) | -2.2712*** (0.7324) | 0.7068 (0.5736) | 4.2922 (3.9609) | -0.2760 (0.8937) |
| Days since SIP | 0.5258*** (0.1158) | 0.2838** (0.1389) | 0.0683 (0.1078) | -0.9193 (0.8325) | 0.0439 (0.1686) |
| Poverty | -0.2059 (1.3737) | 0.2910 (1.7548) | -0.3426 (1.2868) | -4.8845 (12.7673) | -0.5777 (0.5373) |
| Median House Income | -0.0001 (0.0003) | -0.00002 (0.0004) | -0.0002 (0.0003) | -0.0007 (0.0026) | -0.0003*** (0.0001) |
| Lagged Fear Score | -283.7421** (126.2742) | -340.3904** (151.4627) | -326.0898*** (117.4762) | -2,518.8890*** (911.9943) | -139.3457 (183.6979) |
| Political ideology | -48.3021** (22.8208) | -59.4693** (27.5875) | -47.0572** (21.2479) | -440.7139*** (171.0405) | 6.4180 (31.3535) |
| State issued shelter-in-place | -104.5549 (78.8863) | -153.8703 (94.6194) | -149.0849** (73.3898) | -1,364.6140** (569.6519) | -176.8373 (114.9964) |
| Political Ideology X Lagged fear | 300.9612** (119.7042) | 356.3041** (143.5802) | 270.3768** (111.3639) | 2,454.4610*** (864.4575) | 28.6429 (174.2299) |
| SIP X Lagged Fear | 556.4801 (438.4652) | 803.0406 (525.9137) | 792.1204* (407.9147) | 7,504.7270** (3,166.3060) | 932.8997 (639.1718) |
| Political ideology X SIP | 175.6505** (85.3145) | 194.3195* (102.3298) | 174.3164** (79.3701) | 1,799.8810*** (616.0871) | 227.9781* (124.3754) |
| Ideology X SIP X Lagged Fear | -986.9036** (474.1987) | -1,065.4320* (568.7752) | -958.8522** (441.1584) | -9,856.7920*** (3,424.4250) | -1,228.0860* (691.3097) |
| $R^2$ | 0.1035 | 0.1235 | 0.1286 | 0.0313 | 0.1115 |
| $F$ | 175.7882*** | 214.4126*** | 224.6424*** | 49.2411*** | 190.9762*** |

*Note:* $^*p < 0.05$, $^{**}p < 0.01$, $^{***}p < 0.001$

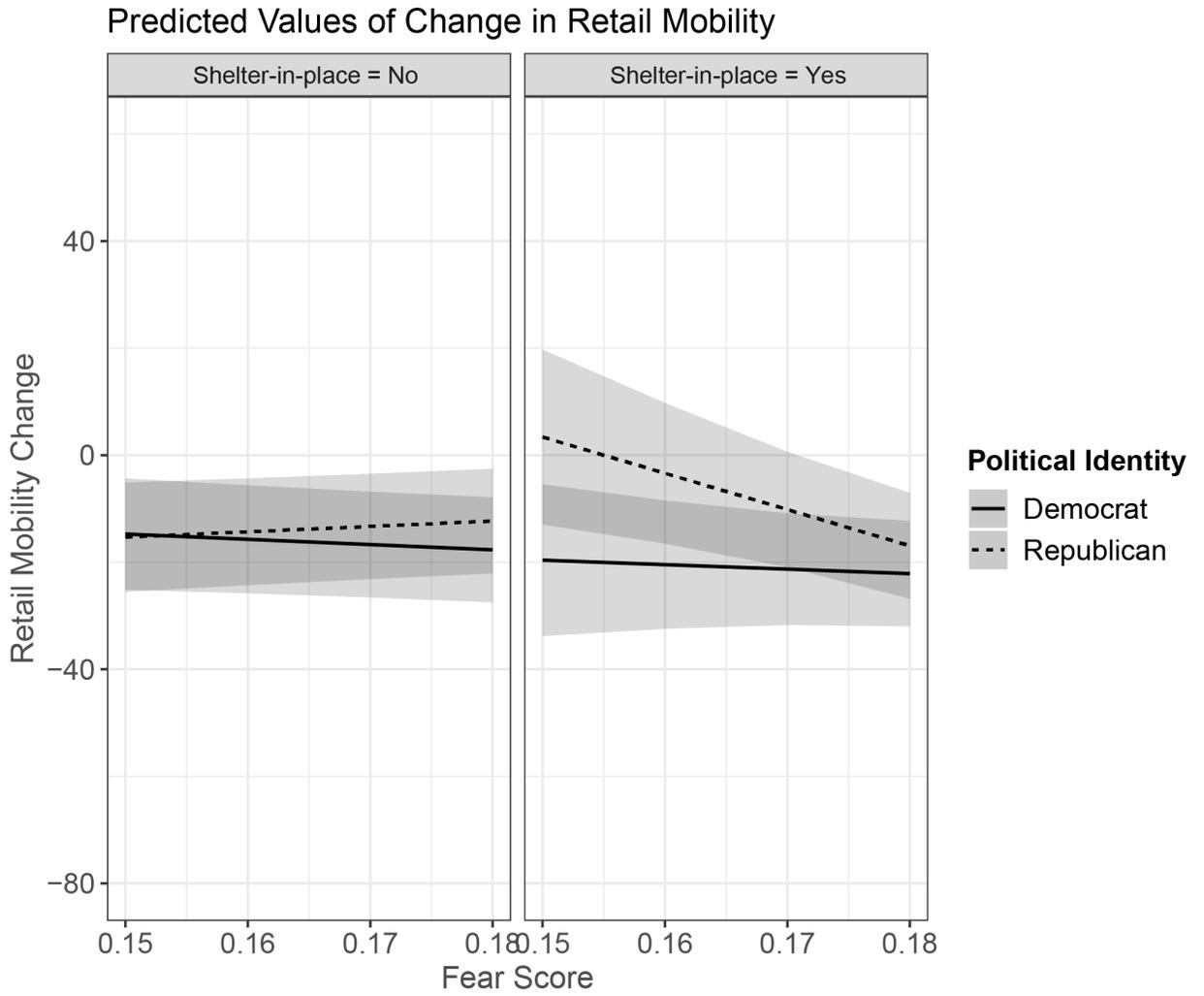

Figure 1: *The marginal effects of the three-way interaction between political voting results, the presence of a shelter-in-place order, and the level of lagged twitter fear sentiment on retail category mobility.*

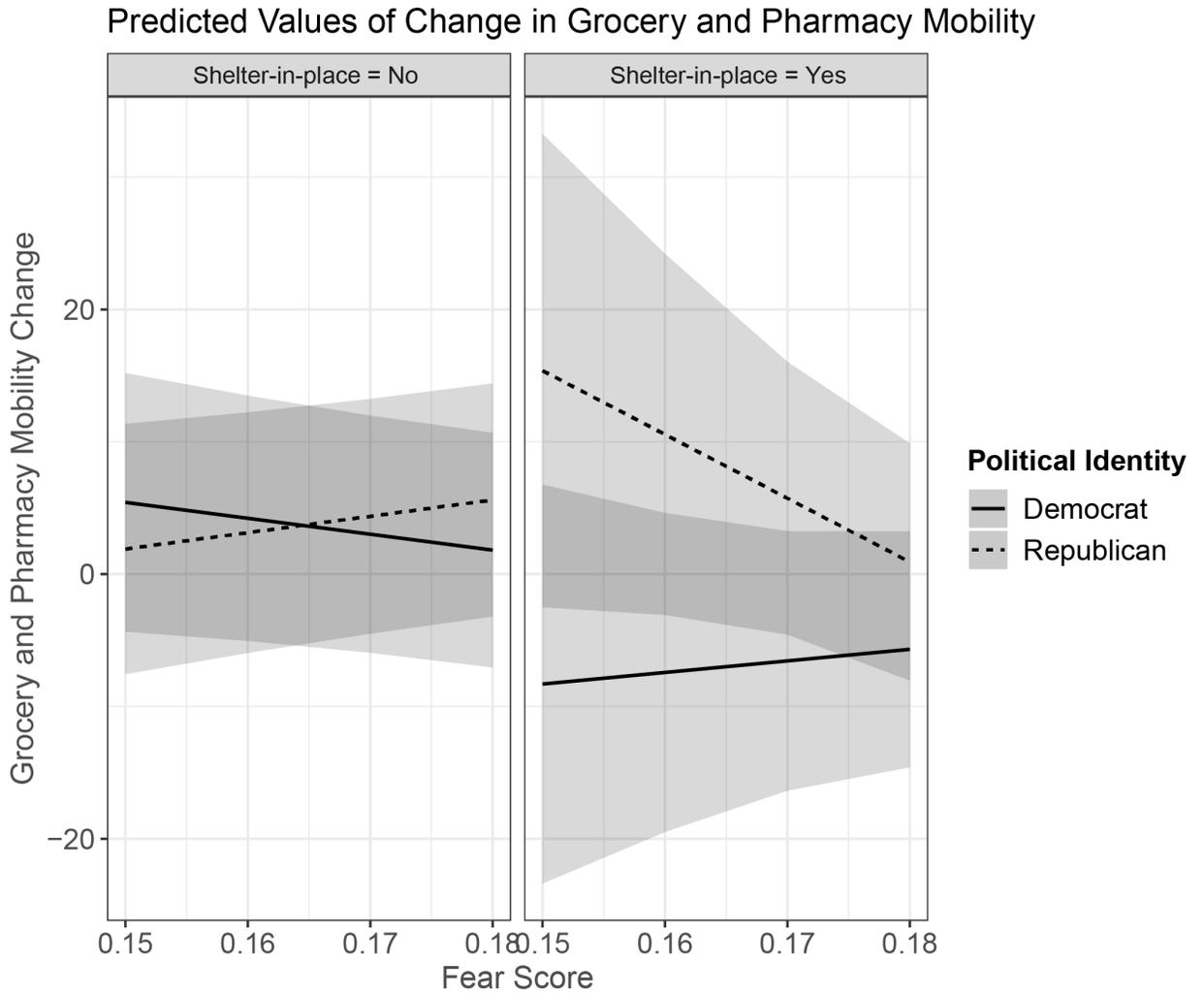

Figure 2: *The three-way interaction explaining grocery category mobility with political voting results, the presence of a shelter-in-place order, and the level of lagged twitter fear sentiment.*

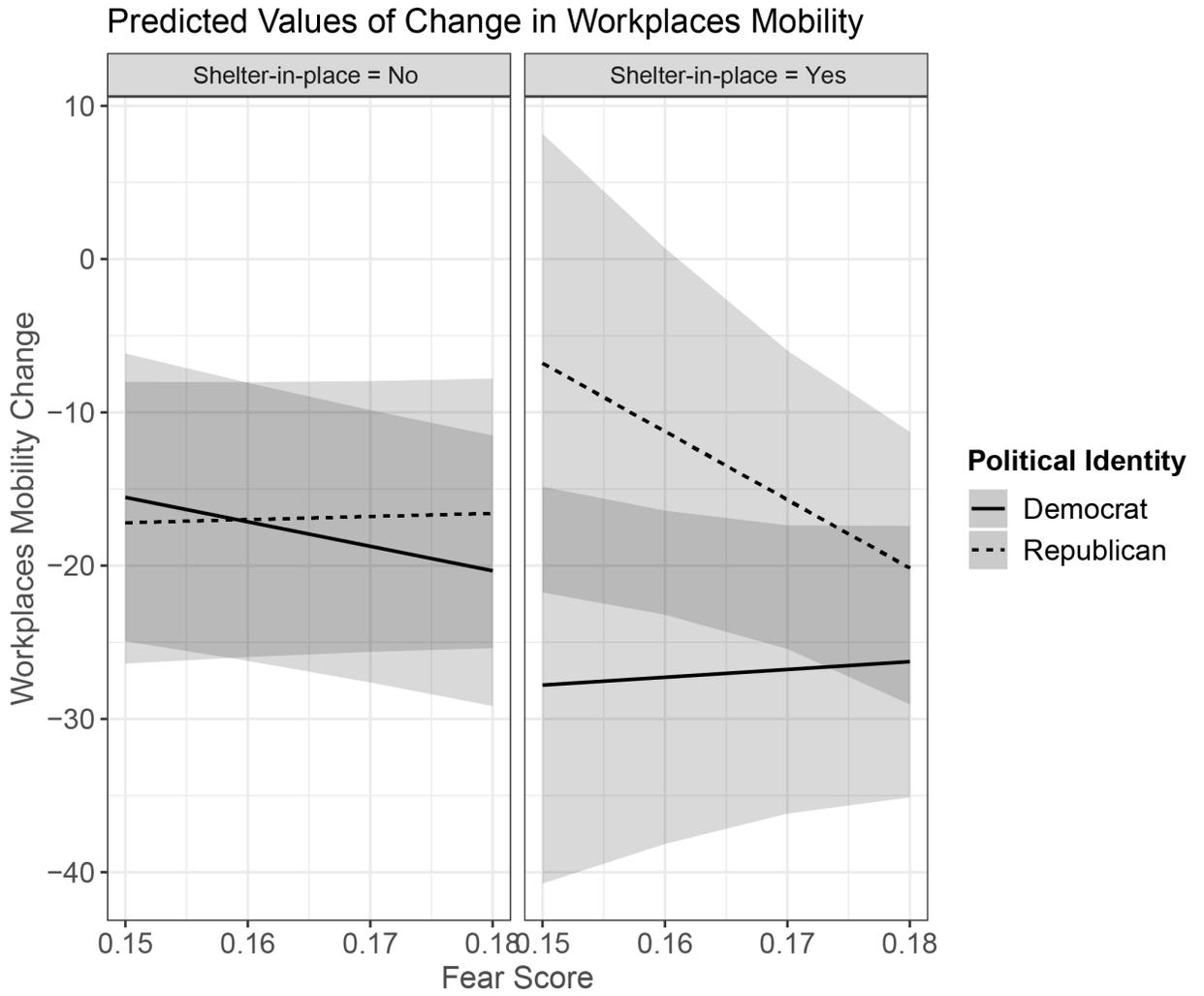

Figure 3: *The three-way interaction explaining workplaces category mobility with political voting results, the presence of a shelter-in-place order, and the level of lagged twitter fear sentiment.*

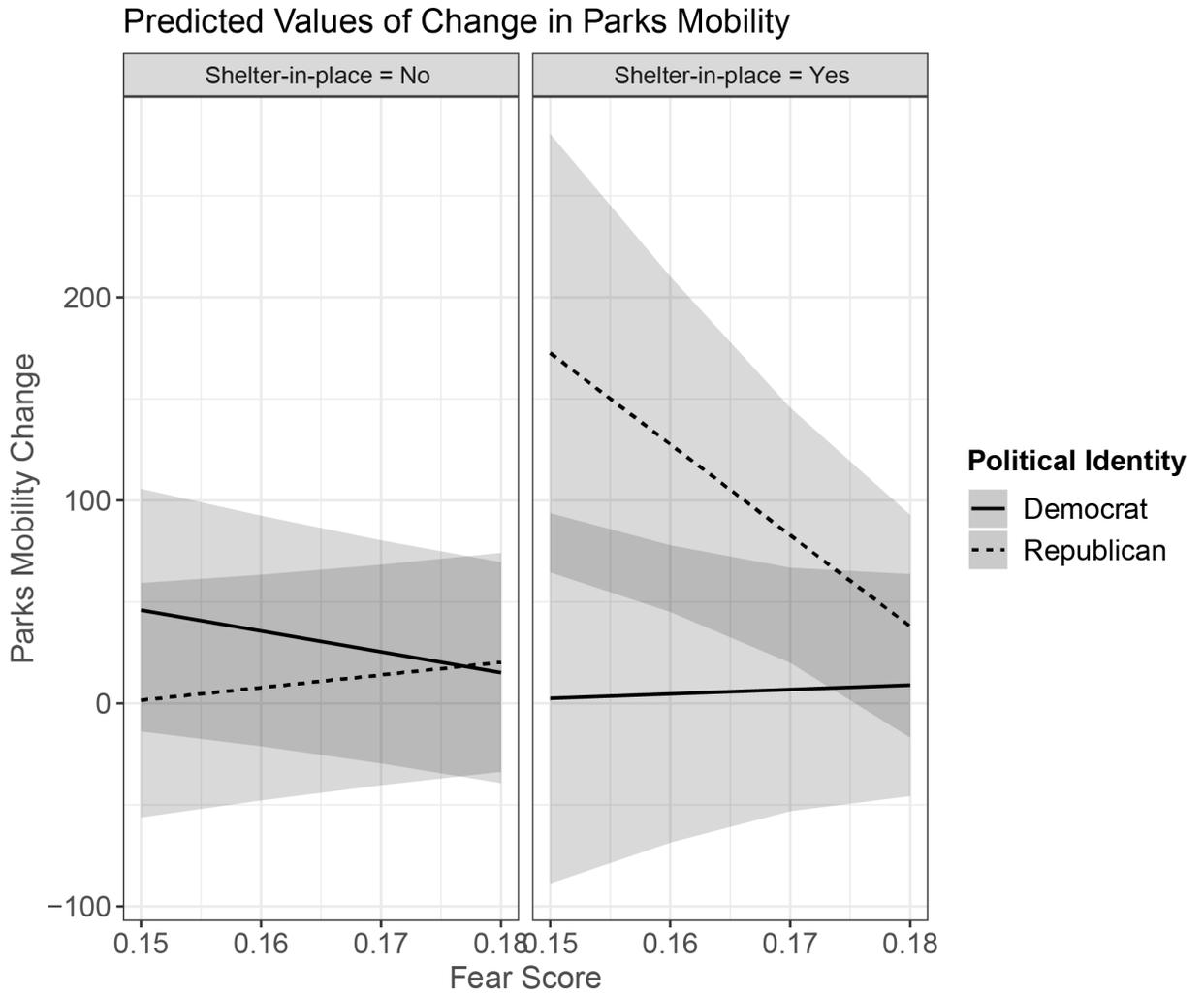

Figure 4: *The three-way interaction explaining parks category mobility with political voting results, the presence of a shelter-in-place order, and the level of lagged twitter fear sentiment.*

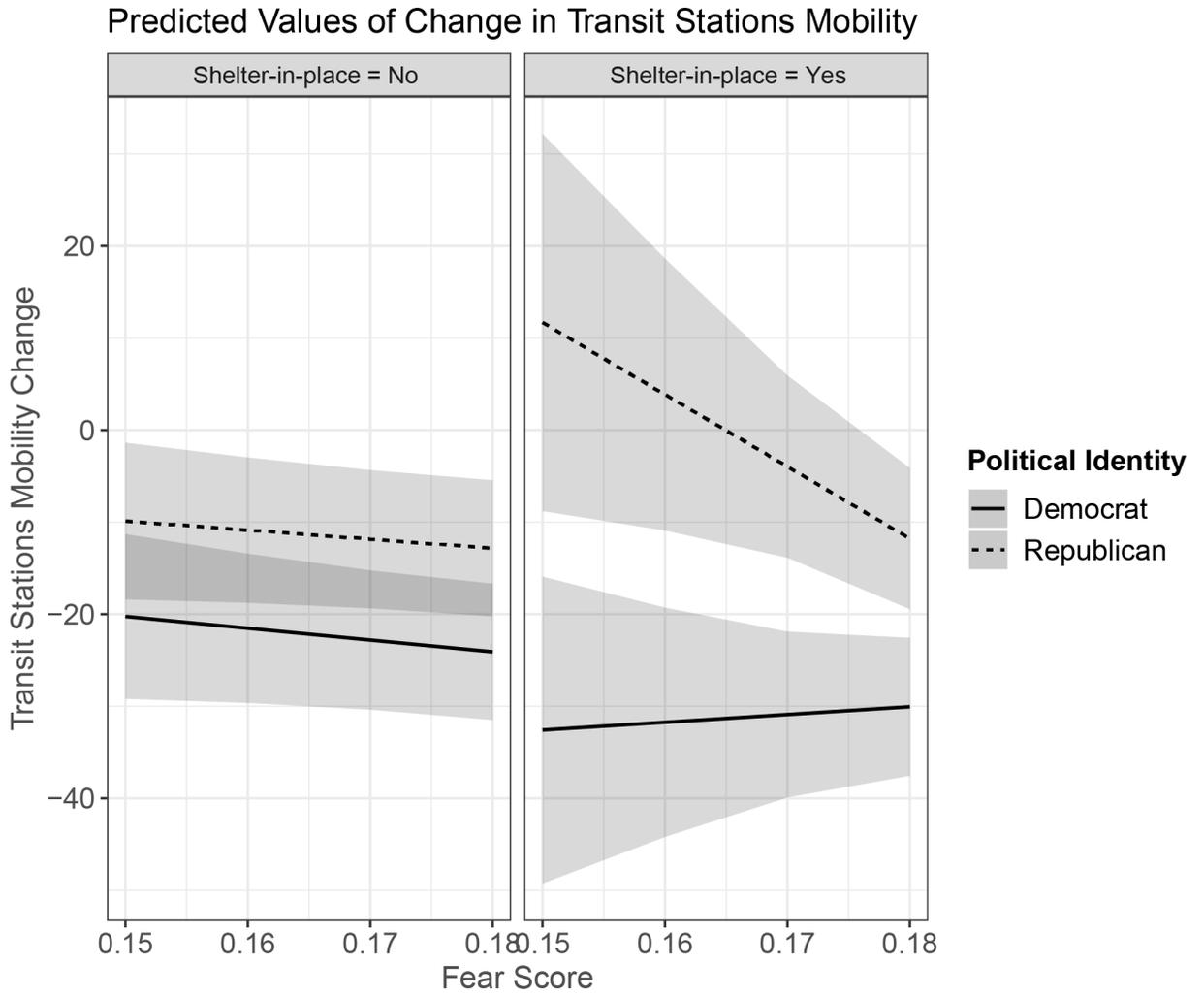

Figure 5: *The three-way interaction explaining transit stations category mobility with political voting results, the presence of a shelter-in-place order, and the level of lagged twitter fear sentiment.*

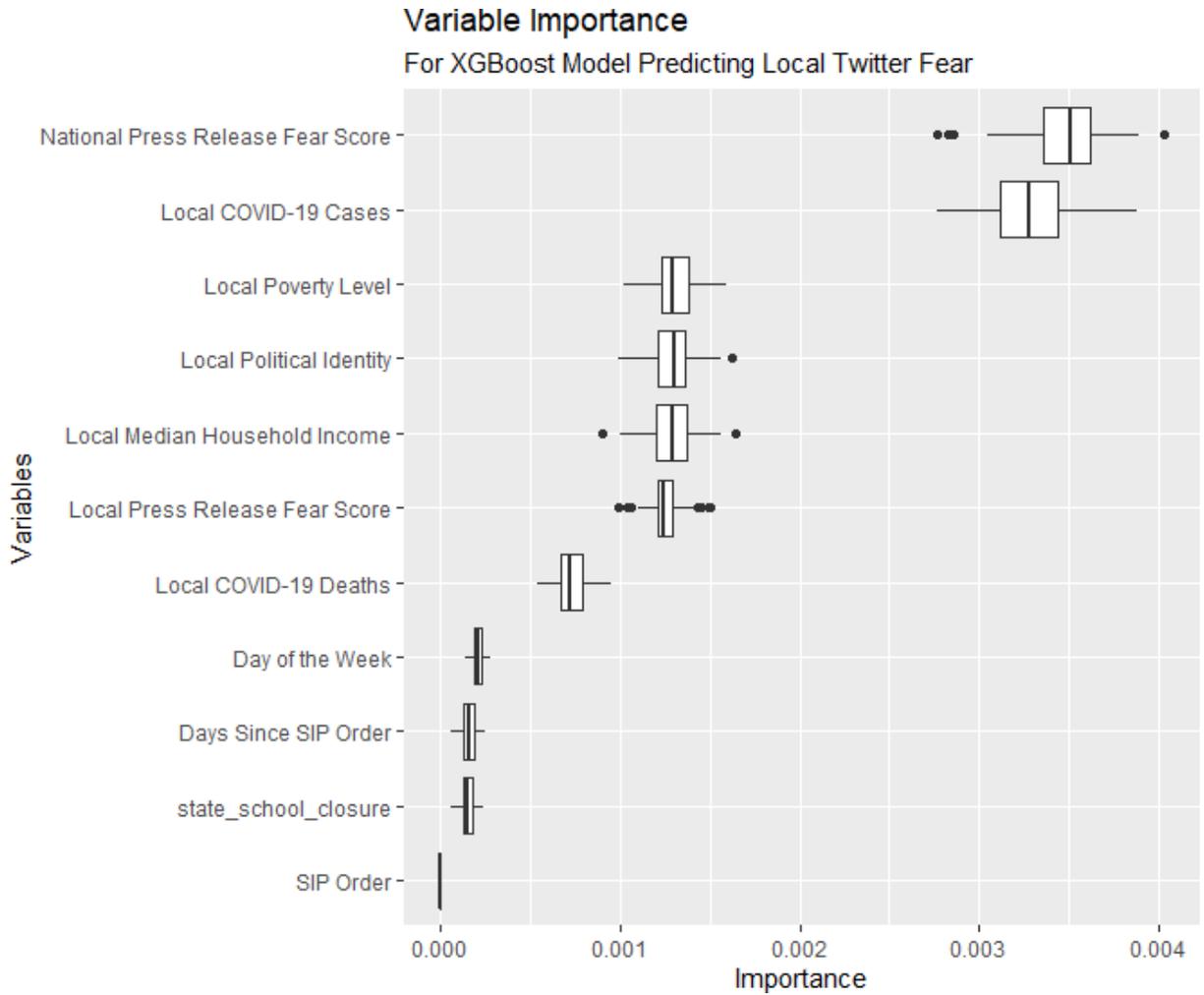

Figure 6: *Variable Importance Plot showing the importance of variables in the XGBoost model predicting Local Tweet Fear. Both National and Local Press Release fear are significant predictors of Local Tweet Fear.*

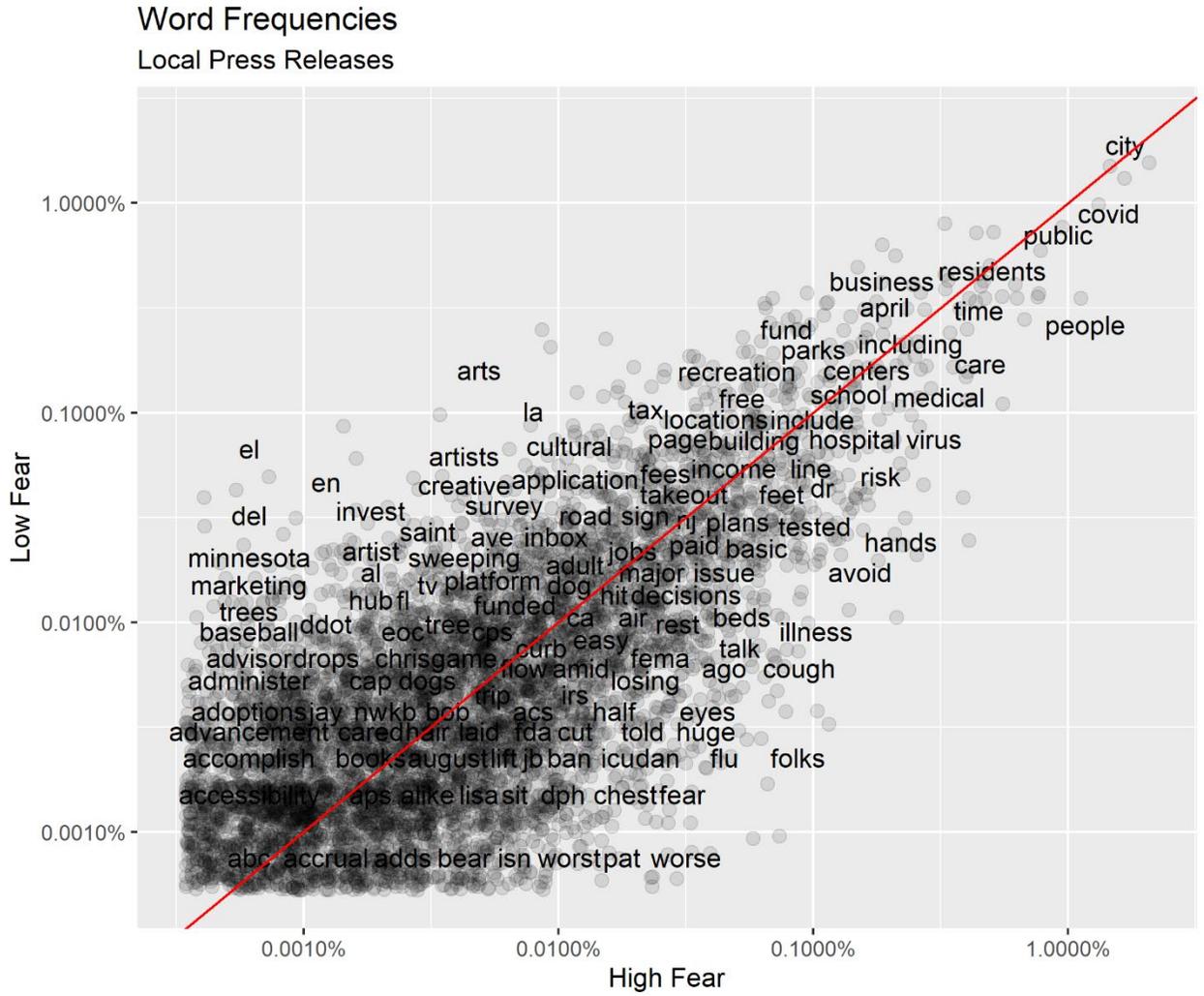

Figure 7: *Plot showing the log odds of individual words between high and low local press releases.*

Figure 8: *Plot showing the log odds of individual words between high and low national press releases.*

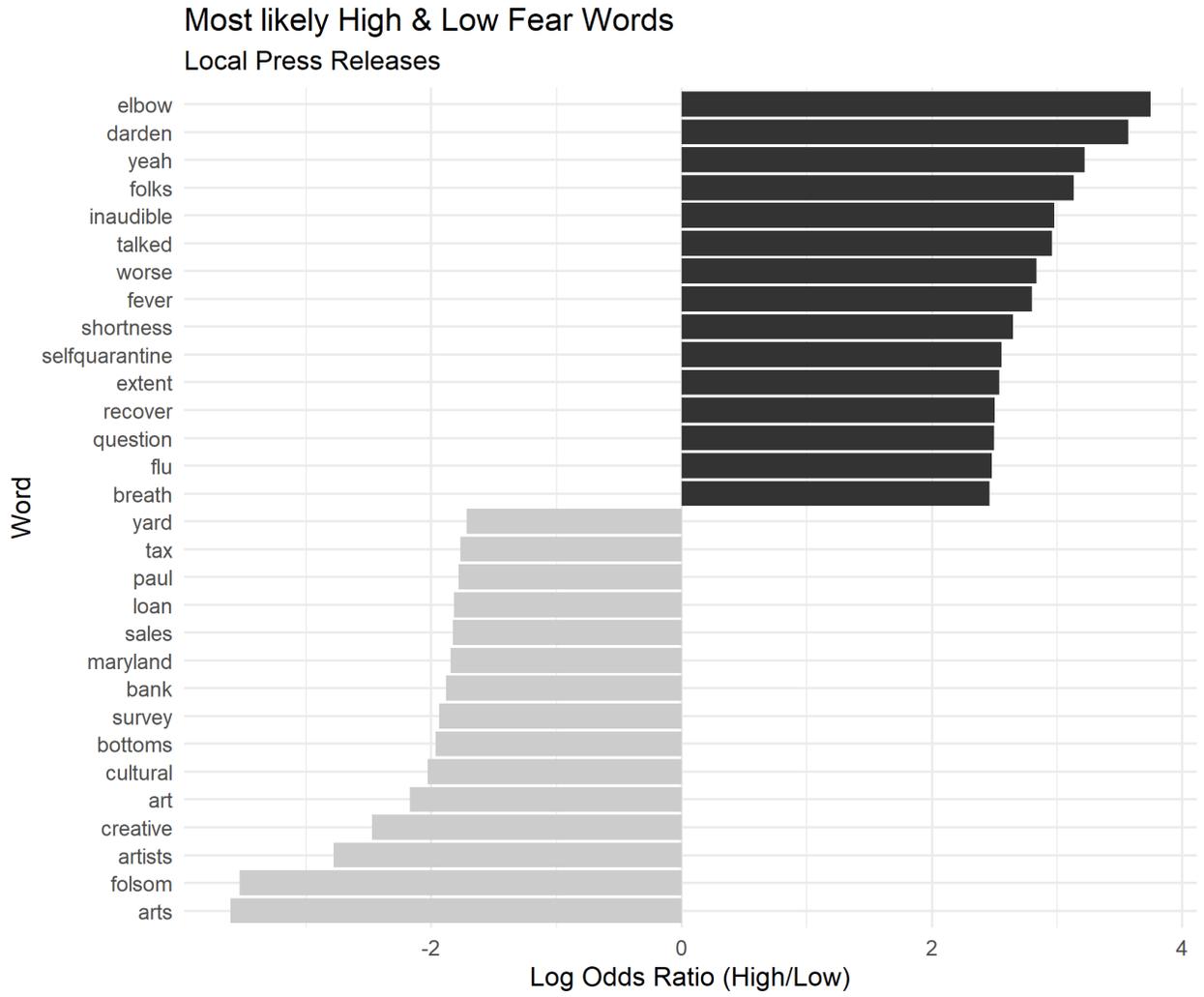

Figure 9: *Plot showing a list of words with the highest likelihood of high and low local press releases.*

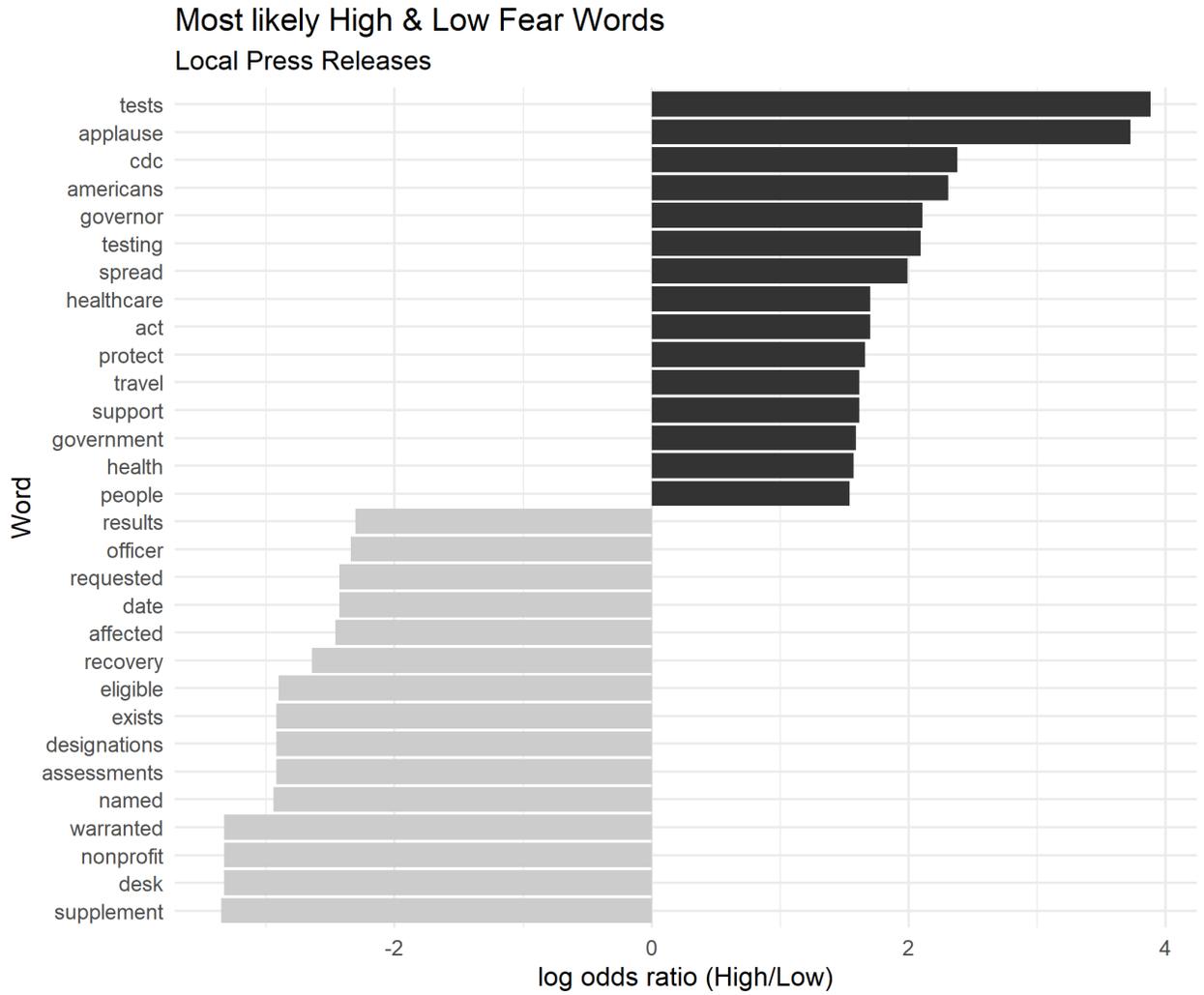

Figure 10: *Plot showing a list of words with the highest likelihood of high and low national press releases.*

## Supplementary Materials

**Appendix A: Data and Variable Explanation**

*Metropolitan Statistical Areas.* Metropolitan statistical areas (metro areas) are designated by the U.S. Census Bureau (2010). The metro areas that had a regional population of more than 1 million residents were selected. All metro areas consist of at least two counties. Any county level data was aggregated to the metro level using a population-weighted mean. Summary statistics are reported in Table 3.

*Shelter-in-place orders.* Any order that focused on restricting movement by penalty was considered a shelter-in-place order, regardless of the name of the order (e.g., "stay at home," "safer at home"). Order dates were collected by searching state and city news sources.

*COVID-19 prevalence.* County-level COVID-19 infection and death rates were collected from Johns Hopkins University (Dong et al., 2020).

*Political Identity.* County level voting data was collected from each presidential election from 2000 through 2016 (MIT, 2019). Country level voting results were aggregated across the five last presidential elections, then aggregated up to the metro level. Using data from the previous five elections controls for shifts and anomalies from the most recent 2016 election. Aggregated election results are calculated as a ratio of votes for the Republican candidate and votes for the Democrat candidate for each metro area. A value greater (less) than one indicates a preference for Republican (Democrat) candidates. In our sample, 17 metropolitan areas are majority Republican voting and 36 are majority Democrat voting.

*Poverty.* County-level poverty rates were collected from the 2018 SAIPE estimate.

Table 3: *A description of all variables used in the studies.*

| | Mean | St. Dev. | Median | Min | Max |
|---|---|---|---|---|---|
| **Google Mobility Categories** | | | | | |
| Retail Mobility | -14.767 | 25.5903 | -7.271 | -66.693 | 24.541 |
| Grocery and Pharmacy Mobility | 2.906 | 14.6828 | 5.419 | -40.075 | 48.579 |
| Workplaces Mobility | -18.372 | 20.9009 | -10.444 | -72.097 | 21.755 |
| Parks Mobility | 18.151 | 40.2986 | 13.777 | -75.210 | 215.423 |
| Transit Stations Mobility | -18.029 | 22.7642 | -11.994 | -74.000 | 28.795 |
| **Main Variables** | | | | | |
| Twitter Fear Sentiment | 0.5572 | 0.0096 | 0.5565 | 0.5109 | 0.6161 |
| Political Identity | 0.9529 | 0.3305 | 0.9107 | 0.3023 | 1.7904 |
| State Ordered SIP | 0.1314 | 0.3380 | 0 | 0 | 1 |
| **Control Variables** | | | | | |
| COVID-19 Confirmed Cases | 76.326 | 259.259 | 5.165 | 0 | 4375.577 |
| COVID-19 Deaths | 1.7734 | 8.2569 | 0 | 0 | 150.0599 |
| Day of the week | 4.002 | 2.0014 | 4 | 1 | 7 |
| State School Closure | 0.4075 | 0.4915 | 0 | 0 | 1 |
| Days Since SIP Order | 0.6786 | 1.9748 | 0 | 0 | 1 |
| Poverty | 11.845 | 2.6494 | 12.105 | 6.297 | 19.291 |
| Median Household Income | 69985 | 14639.57 | 66135 | 50697 | 124553 |

## Appendix B: Metropolitan Tweet Decision Rules

Metro areas were collected from the optional location field on each Twitter user's profile.

Individual tweets were assigned to a metropolitan area based on a string search of the location field. Search terms used to select users for each metropolitan area are listed in Table 4.

Table 4: *A list of all metropolitan areas and the search strings used to identify a local area within the Twitter bio.*

| Metropolitan Area | Search Strings |
|---|---|
| New York-Newark-Jersey City, NY-NJ-PA | nyc, new york city, newark, jersey city |
| Dallas-Fort Worth-Arlington, TX | dallas, fort worth |
| Houston-The Woodlands-Sugar Land, TX | houston, the woodlands, sugar land |
| Boston-Cambridge-Newton, MA-NH | boston, cambridge, newton |
| Phoenix-Mesa-Chandler, AZ | phoenix, mesa, chandler |
| San Francisco-Oakland-Berkeley, CA | san francisco, sf, oakland, berkeley |
| Detroit-Warren-Dearborn, MI | detroit, warren, dearborn |
| Seattle-Tacoma-Bellevue, WA | seattle, tacoma, bellevue |
| Minneapolis-St. Paul-Bloomington, MN-WI | minneapolis, st paul, st. paul, bloomington |
| Tampa-St. Petersburg-Clearwater, FL | tampa, petersburg, clearwater |
| Denver-Aurora-Lakewood, CO | denver, aurora, lakewood |
| St. Louis, MO-IL | st louis, st. louis |
| Baltimore-Columbia-Towson, MD | baltimore, columbia, towson |
| Orlando-Kissimmee-Sanford, FL | orlando, kissimmee, sanford |
| Charlotte-Concord-Gastonia, NC-SC | charlotte, concord, gastonia |
| San Antonio-New Braunfels, TX | san antonio, new braunfels |
| Portland-Vancouver-Hillsboro, OR-WA | portland, hillsboro |
| Sacramento–Roseville–Folsom, CA | sacramento, roseville, folsom |
| Pittsburgh, PA | pittsburgh |
| Las Vegas-Henderson-Paradise, NV | vegas, henderson |
| Cincinnati, OH-KY-IN | cincinnati |
| Austin-Round Rock-Georgetown, TX | austin, round rock, georgetown |
| Kansas City, MO-KS | kansas city |
| Columbus, OH | columbus |
| Cleveland-Elyria, OH | cleveland, elyria |
| Indianapolis-Carmel-Anderson, IN | indianapolis, carmel, anderson |
| San Jose-Sunnyvale-Santa Clara, CA | sj, san jose, sunnyvale, santa clara |
| Milwaukee-Waukesha, WI | milwaukee, waukesha |
| Jacksonville, FL | jacksonville |
| Oklahoma City, OK | oklahoma city |
| Raleigh-Cary, NC | raleigh, cary |
| Memphis, TN-MS-AR | memphis |
| Richmond, VA | richmond |
| Louisville/Jefferson County, KY-IN | louisvilla, jefferson county |
| New Orleans-Metairie, LA | new orleans, metairie |
| Salt Lake City, UT | slc, salt lake city |
| Hartford-East Hartford-Middletown, CT | hartford, middletown |
| Buffalo-Cheektowaga, NY | buffalo, cheektowaga |
| Rochester, NY | rochester |
| Grand Rapids-Kentwood, MI | grand rapids, kentwood |
| Tucson, AZ | tucson |

## Appendix C: Dictionary Method Selection

We initially compared the word vectors of 35 synonyms of fear to the word vector for fear (Nicolas et al., 2019), calculated the cosine similarity, created two distinct clusters of words, and created a dictionary from the tightest cluster to use in the previously described distributed dictionary representation (DDR) method (Garten et al., 2018). The final validated dictionary consists of the 26 words: abhor, alarm, anxiety, apprehension, aversion, concern, consternation, cowardice, doubt, dread, fear, fearful, foreboding, fright, horror, nerves, nervous, nightmare, panic, perturb, scare, terror, trepidation, unease, unrest, worry. The cosine similarity of the initial 35 synonyms to the word fear can been seen in table 5. Clusters can be seen in figure 11.

Figure 11: *Clusters of words based on cosine similarity to the word 'fear'. This method created a dictionary of 26 words (cluster 1) most similar to the word 'fear'.*

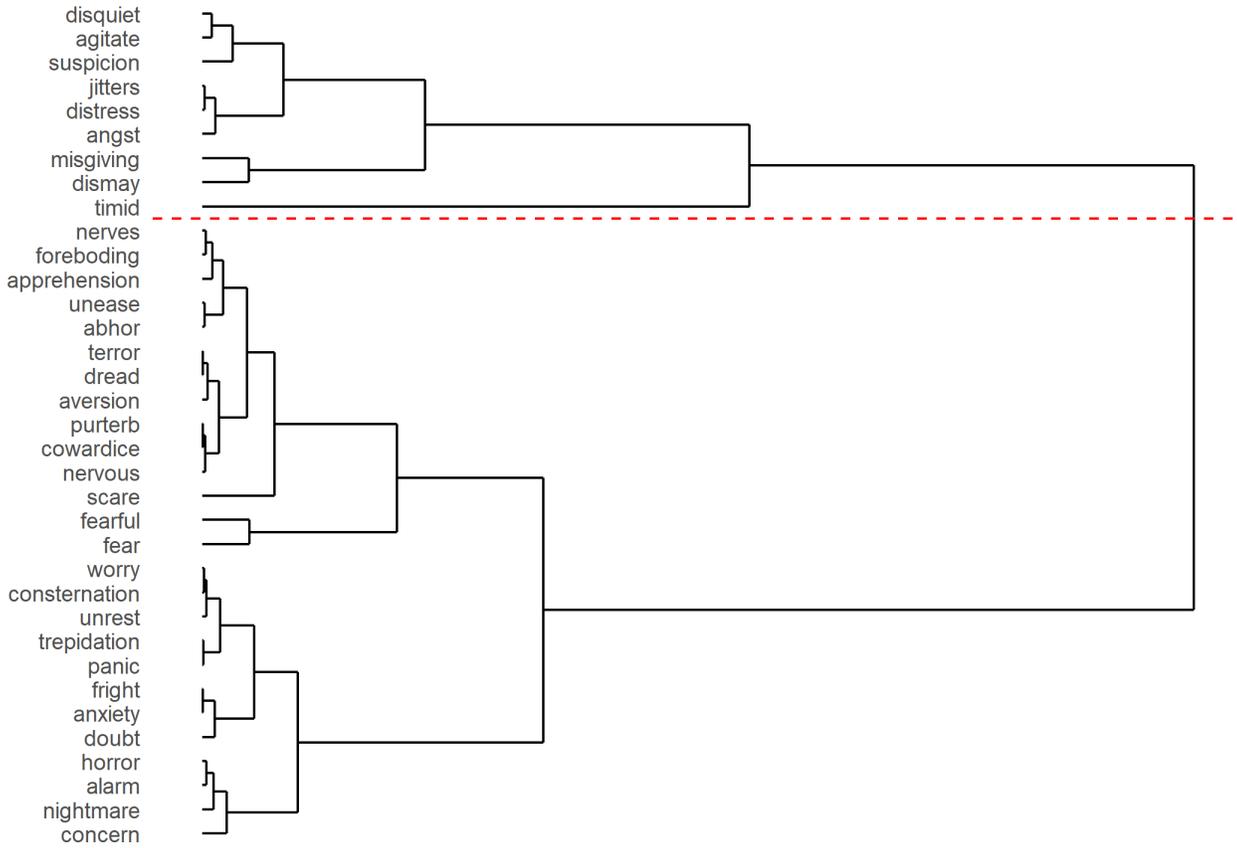

Table 5: *A table showing the cosine similarity results for the 35 collected synonyms of the word fear.*

| Synonym | Cosine Similarity |
| --- | --- |
| fear | 1 |
| fearful | 0.985727 |
| scare | 0.962931 |
| abhor | 0.954734 |
| unease | 0.954072 |
| foreboding | 0.951454 |
| nerves | 0.950261 |
| apprehension | 0.948438 |
| terror | 0.946125 |
| dread | 0.946003 |
| aversion | 0.944522 |
| perturb | 0.941986 |
| cowardice | 0.94175 |
| nervous | 0.941088 |
| concern | 0.925903 |
| horror | 0.921962 |
| alarm | 0.920626 |
| nightmare | 0.918454 |
| unrest | 0.912737 |
| worry | 0.912082 |
| consternation | 0.911427 |
| trepidation | 0.907766 |
| panic | 0.907329 |
| anxiety | 0.900682 |
| fright | 0.900566 |
| doubt | 0.896939 |
| suspicion | 0.865655 |
| agitate | 0.859307 |
| disquiet | 0.856447 |
| angst | 0.84493 |
| jitters | 0.841719 |
| distress | 0.841007 |
| dismay | 0.812474 |
| misgiving | 0.798335 |
| timid | 0.700273 |

**Appendix D: Lagged Variable Diagnostic Results**

As stated in the main discussion, standard diagnostics for selecting a lag structure recommended a lag value of three for all mobility categories. We chose to implement a lag value of one because of the unique nature of Twitter and tweet impact (Romero et al., 2011; Ye & Wu,

2010). The AIC and BIC results for each mobility category and lag value can be seen in table 6.

Table 6: *The AIC and BIC scores from random effects models with combinations of mobility categories and fear lag variables. Results show a preference for a lag value of 3.*

| Mobility Category | Twitter Fear Lag | AIC | BIC |
|---|---|---|---|
| Retail | 1 | 8815.614 | 8895.678 |
| | 2 | 8486.334 | 8565.871 |
| | 3 | **8210.736** | **8289.728** |
| Grocery and Pharmacy | 1 | 9374.76 | 9454.823 |
| | 2 | 9071.618 | 9151.155 |
| | 3 | **8776.064** | **8855.056** |
| Workplaces | 1 | 8593.603 | 8673.666 |
| | 2 | 8297.556 | 8377.094 |
| | 3 | **8009.553** | **8088.545** |
| Parks | 1 | 14894.79 | 14974.85 |
| | 2 | 14392.73 | 14472.27 |
| | 3 | **13878.42** | **13957.42** |
| Transit Stations | 1 | 9976.949 | 10057.01 |
| | 2 | 9584.285 | 9663.822 |
| | 3 | **9217.166** | **9296.158** |

## Appendix E: Model Selection Diagnostic Results

The F-test, Breusch-Pagan LaGrange Multiplier test, and Hausman test are used to compare model fit between fully pooled (OLS), fixed effects, and random effects models. Instead of showing results for each model, we show the results from the diagnostic tests which indicate that the random effects model is preferred for all mobility categories. An F-test p-value below 0.05 (true for all mobility categories) indicates that the fixed effects model is preferred over the fully pooled model. A LaGrange Multiplier p-value below 0.05 (true for all mobility categories) indicates that the random effects model is preferred. Finally, a Hausman test p-value below 0.05 (not true for any mobility categories) would indicate that the fixed effects model is preferred over the random effects model. From these results, the final model selected was the random effects model for all mobility categories.

Table 7: *A table showing the F test, Breusch-Pagan LaGrange Multiplier test, and Hausman test for each of the mobility category models which include all variables shown*

*in the paper. The results show that the random effects models are preferred for all mobility category models.*

| Mobility Category | F Test | Breusch-Pagan LaGrange Multiplier Test | Hausman Test |
|---|---|---|---|
| | **p-value** | **p-value** | **p-value** |
| Retail | < 0.0001 | < 0.0001 | 0.9689 |
| Grocery | < 0.0001 | < 0.0001 | 0.7822 |
| Workplaces | < 0.0001 | < 0.0001 | 0.9632 |
| Parks | < 0.0001 | < 0.0001 | 0.4325 |
| Transit Stations | < 0.0001 | < 0.0001 | 0.0659 |